\documentclass[prl,aps,twocolumn,floatfix,showpacs]{revtex4}
\usepackage{graphicx,rotating,subfigure,amsmath,amsfonts,amssymb,delarray}
\usepackage{dsfont}
\usepackage[T1]{fontenc}
\renewcommand{\vec}[1]{\boldsymbol #1}
\newcommand{\e}{\text{e}}

\def\12{\frac{1}{2}}

\newcommand{\be}{\begin{equation}}
\newcommand{\ee}{\end{equation}}
\newcommand{\bea}{\begin{eqnarray}}
\newcommand{\eea}{\end{eqnarray}}

\DeclareMathOperator\Tr{Tr}

\begin{document}
\bibliographystyle{apsrev}
\title{Entanglement measures and the quantum to classical mapping} 


\author{Jesko Sirker} 
\affiliation{Department of Physics and Research Center OPTIMAS,
  Technical University Kaiserslautern,
  D-67663 Kaiserslautern, Germany} 
\date{\today}

\begin{abstract}
  A quantum model can be mapped to a classical model in one higher
  dimension. Here we introduce a finite-temperature correlation
  measure based on a reduced density matrix $\bar\rho_{\bar A}$
  obtained by cutting the classical system along the imaginary time
  (inverse temperature) axis. We show that the von-Neumann entropy
  $\bar S_{\rm ent}$ of $\bar\rho_{\bar A}$ shares many properties
  with the mutual information, yet is based on a simpler geometry and
  is thus easier to calculate. For one-dimensional quantum systems in
  the thermodynamic limit we proof that $\bar S_{\rm ent}$ is
  non-extensive for all temperatures $T$. For the integrable
  transverse Ising and $XXZ$ models we demonstrate that the
  entanglement spectra of $\bar\rho_{\bar A}$ in the limit $T\to 0$
  are described by free-fermion Hamiltonians and reduce to those of
  the regular reduced density matrix $\rho_A$---obtained by a spatial
  instead of an imaginary-time cut---up to degeneracies.
\end{abstract}


\maketitle
Entanglement is usually thought of as a quantum mechanical entity yet
it is well known that the properties of quantum models can be computed
from a classical model in one dimension higher
\cite{Suzuki1b,Zinn-Justin,Sachdev}. Such an approach is particularly
useful if one is interested in how entanglement builds up in a thermal
state while lowering the temperature or during unitary time evolution.

Entanglement measures are, generally speaking, maps from the space of
density matrices into the positive real numbers. Various entanglement
measures have been introduced in the last decades and a set of axioms
has been put forward which a good measure should possess
\cite{VedralPlenio,PlenioVirmani}.
For bipartite pure states the von-Neumann entropy of entanglement
fulfills these criteria and provides a bridge into statistical
mechanics and condensed matter physics. In particular, the reduced
density matrix, whose eigenvalues determine the entanglement entropy,
is at the heart of the density matrix renormalization group (DMRG)
\cite{WhiteDMRG}. This numerical method allows one to optimally
approximate pure states of many-body systems by matrix product states
\cite{OestlundRommer} and is most successful in one dimension.

In experiments we are, however, usually dealing with mixed states
about whose entanglement properties much less is known. Entanglement
measures commonly used for pure states of multi-particle systems such
as the entanglement entropy fail because they become extensive and
thus no longer fulfill a boundary law. Furthermore, they cannot
distinguish between classical and quantum correlations
\cite{VedralPlenio}. Entanglement measures which do distinguish
between these different types of correlations 
involve extremizations over all possible decompositions of the density
matrix and explicit results have only been obtained for few-particle
density matrices \cite{Wootters,ArnesenBose,Dillenschneider}. Putting
these fundamental difficulties aside, it is still useful to define
correlation measures for thermal ensembles which go beyond the one-
and two-point correlation functions traditionally studied in
statistical mechanics and condensed matter physics. They might help to
reveal, in particular, phase transitions with complex or topological
order parameters.

In this letter we want to investigate how correlations, generated
during imaginary time evolution, can be quantified. After a quantum to
classical mapping we introduce as entanglement measure the von-Neumann
entropy of a reduced density matrix obtained by a partial trace in the
imaginary time direction. This measure also cannot distinguish between
quantum and classical correlations but we will show that it shares
many properties with the mutual information which has recently
attracted considerable interest as a correlation witness for many-body
systems
\cite{IsakovHastings,SinghHastings,WilmsVidal,WilmsTroyer}. The
measure introduced in this letter is also a natural choice from the
perspective of numerical matrix product state algorithms: While the
reduced density matrix obtained by a spatial trace is at the heart of
the DMRG at $T=0$, the reduced density matrix considered here is used
in transfer-matrix DMRG algorithms to study the finite-temperature
properties of one-dimensional systems in the thermodynamic limit
\cite{BursillXiang,WangXiang,Shibata,SirkerKluemperEPL,SirkerFidelity}.

The entanglement entropy for a bipartite system, $S=A\cup B$, is
defined as
\begin{equation}
\label{Sent}
S_{\rm ent}(A)=-\Tr\rho_A\ln\rho_A
\end{equation}
where $\rho_A=\Tr_B \rho$, $\Tr\rho_A=1$, is a reduced density matrix
obtained from the density matrix $\rho$ of the system $S$
by spatially tracing out part $B$. If the system is in a pure state,
then it is easy to show by a Schmidt decomposition that $S_{\rm
  ent}\equiv S_{\rm ent}(A)= S_{\rm ent}(B)$. It follows that $S_{\rm
  ent}$ cannot be extensive but rather has to scale with the surface
between regions $A$ and $B$ \cite{Srednicki_Sent}. In critical
one-dimensional systems this boundary law is known to be weakly
violated by logarithmic corrections
\cite{HolzheyLarsen,CalabreseCardy}.
Here we want to study the mixed state described by the canonical
density matrix $\rho_{\rm c}=\exp(-\beta H)/Z$ where $\beta$ is the
inverse temperature, $H$ the Hamiltonian, and $Z=\Tr\rho_{\rm c}$ the
partition function. Calculating the entanglement entropy (\ref{Sent})
for $\rho_c$ one finds that $S_{\rm ent}(A)\neq S_{\rm ent}(B)$ in
general and that $S_{\rm ent}$ becomes an {\it extensive} quantity
approaching the regular thermal von-Neumann entropy for $\beta\to 0$
\cite{CalabreseCardy,SorensenMingChyang}.  This can be easily
understood as follows: If all correlation lengths are much smaller
than the extent of the considered subsystem then the rest of the
system just acts as an additional bath.
A way to partly correct this is to consider
the {mutual information} given by $\mathcal{I}_{A,B}=S_{\rm
  ent}(A)+S_{\rm ent}(B)-S_{\rm th}(A\cup B)$ where $S_{\rm th}(A\cup
B)=-\Tr\rho_c\ln\rho_c$ is the von-Neumann entropy of the whole
system.
The thermal contribution is then explicitly subtracted ensuring that
$\mathcal{I}_{A,B}(\beta\to 0)\to 0$ as required.  However, the mutual
information is still not a proper entanglement measure because
classical as well as quantum correlations contribute. Furthermore, the
evaluation of the mutual information even when using Renyi instead of
von-Neumann entropies is quite involved due to the complicated
geometry required to obtain generalized partition functions
\cite{CalabreseCardy,MelkoKallin,IsakovHastings,SinghHastings,WilmsTroyer}.
An obvious question is if related finite-temperature correlation
measures can be defined which are easier to use in analytical and
numerical calculations.

\begin{figure}[!ht]
\begin{center}
\includegraphics*[width=0.98\columnwidth]{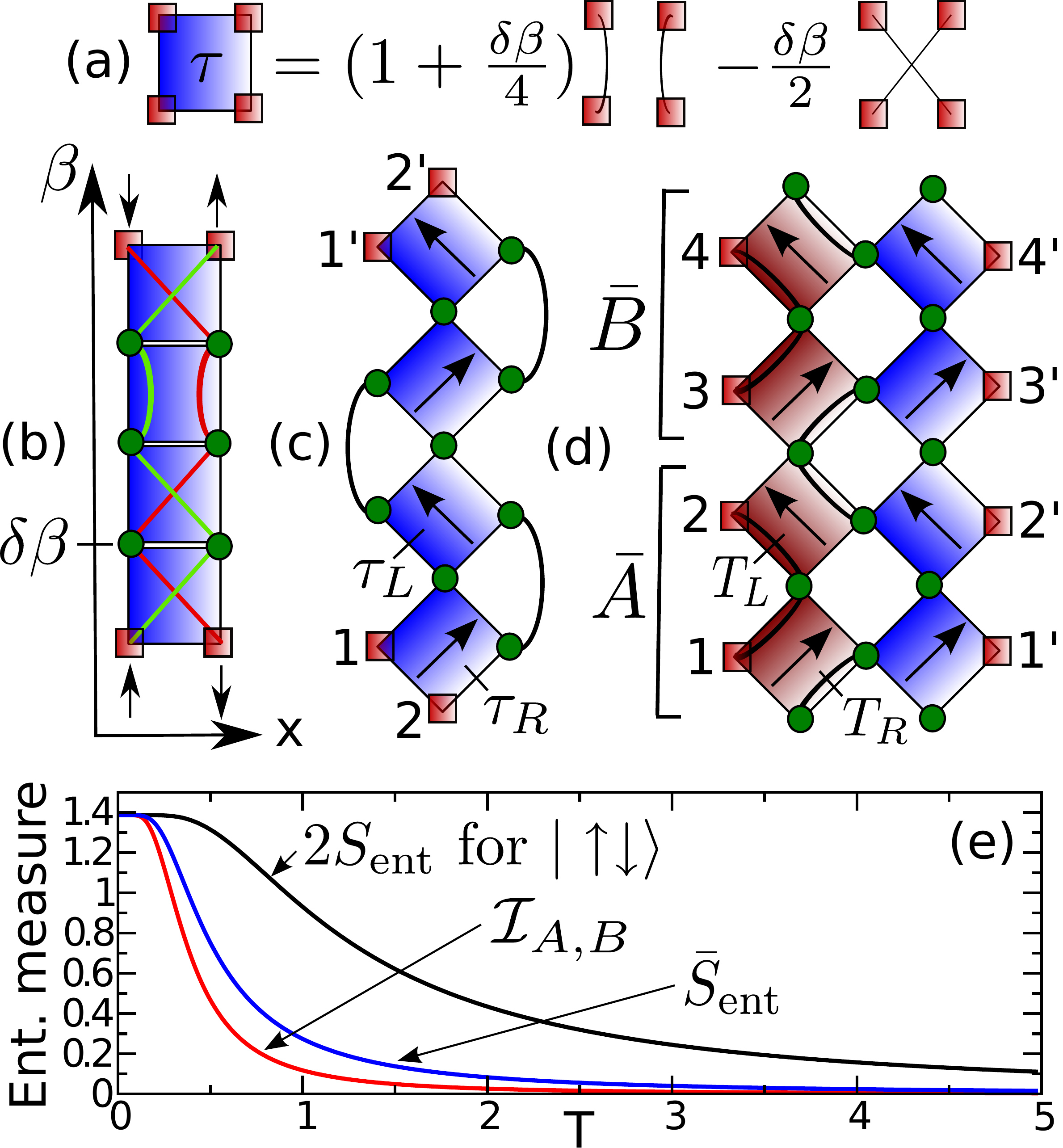}
\end{center}
\caption{(Color online) (a) Local Boltzmann weight $\tau$,
  Eq.~(\ref{tau}). 
  (b) Braiding of worldlines during imaginary time evolution
  discretized in steps $\delta\beta$ with green circles denoting
  tensor contractions. (c) Density matrix $\rho_c(1,2|1',2')$, the
  lines connect contracted indices. (d) Density matrix
  $\bar\rho(1,2,3,4|1',2',3',4')$ projecting onto the ground state in
  auxiliary space. (e) Entanglement measures as function of
  temperature.}
\label{Fig1}
\end{figure}

\paragraph*{Coupled qubits}
To motivate the idea of a correlation measure based on a
quantum-to-classical mapping we start with two coupled qubits with
Hamiltonian $H=\vec{S}_1\cdot\vec{S}_2$ where $\vec{S}$ is a
spin-$1/2$ operator. Let us consider first the imaginary time
evolution starting from the separable state
$\mid\uparrow\downarrow\rangle =(\mid\uparrow\downarrow\rangle +
\mid\downarrow\uparrow\rangle)/2+(\mid\uparrow\downarrow\rangle -
\mid\downarrow\uparrow\rangle)/2$. Time evolving with $\e^{-\beta H}$
this state becomes $|\Psi_\beta\rangle =
\frac{\exp(-\beta/4)}{2}(\mid\uparrow\downarrow\rangle +
\mid\downarrow\uparrow\rangle)+\frac{\exp(3\beta/4)}{2}(\mid\uparrow\downarrow\rangle
- \mid\downarrow\uparrow\rangle)$ and gets projected onto the
maximally entangled singlet ground state for $\beta\to\infty$. A
picture showing how the entangled state is formed out of the separable
state 
can be obtained by the quantum-to-classical mapping. We discretize
time into steps $\delta\beta$ writing $\rho_{\rm c}=\tau^M/Z$ where
$\tau=\exp(-\delta\beta\, H)$ and $M\delta\beta=\beta$. Rewriting the
Hamiltonian as $H=\frac{1}{2}P_{1,2}-\frac{1}{4}\mbox{Id}_{1,2}$ we
find
\begin{equation}
\label{tau}
\tau \approx (1+\frac{\delta\beta}{4})\mbox{Id}_{1,2} -\frac{\delta\beta}{2} P_{1,2}
\end{equation}
where $\mbox{Id}_{1,2}$ is the identity and $P_{1,2}$ the operator
permuting the spins at sites $1,2$ during the time step $\delta\beta$.
Entanglement is thus being generated during imaginary time evolution
by the braiding of the worldlines of the two spins. This is shown
pictorially for one possible configuration of $\tau$-matrices in
Fig.~\ref{Fig1}(a). For the separable initial state
$\mid\uparrow\downarrow\rangle$, the correlations created can be
measured by $S_{\rm ent}$, Eq.~(\ref{Sent}), using
$\rho=|\Psi_\beta\rangle\langle\Psi_\beta|/\langle\Psi_\beta|\Psi_\beta\rangle$
as density matrix and tracing out one of the spins (see
Fig.~\ref{Fig1}(e)).
When using $\rho_{\rm c}$ as the density matrix
, on the other hand, one finds $S_{\rm ent}=\ln 2$ independent of
temperature. Thus $S_{\rm ent}$ fails as a correlation measure and the
whole temperature dependence of the mutual information, see
Fig.~\ref{Fig1}(e), stems from the thermal von-Neumann entropy with
$\mathcal{I}_{A,B}\sim 3/32T^2$ for $T\to\infty$.

Here we want to pursue a different perspective on the classical
representation of the qubits shown in Fig.~\ref{Fig1}(b) by defining a
transfer matrix operator, a method often used in statistical
mechanics.
One way to achieve this is to perform an alternating $45^\circ$
clockwise and anti-clockwise rotation of the $\tau$-plaquettes
starting from Fig.~\ref{Fig1}(b). This leads to the completely
equivalent graphical representation shown in Fig.~\ref{Fig1}(c) with
plaquettes $\tau_{R,L}=T_{R,L}\tau$ where $T_{R,L}$ is the right/left
shift operator, respectively. 
Tracing either over the pair of open indices $1,1'$ or $2,2'$ gives
the reduced density matrix $\rho_A$ while tracing over both pairs
gives the partition function $Z$. By moving the right $\delta$ bonds
(lines in Fig.~\ref{Fig1}(c)) to the left and tracing over $1,1'$ and
$2,2'$ we obtain the density matrix $\bar\rho$ acting in auxiliary
space along the imaginary time axis. $\bar\rho$, shown in
Fig.~\ref{Fig1}(d), consists of two columns with the left column
containing only right and left shift operators, $T_{R,L}$.
Contracting pairwise the open indices $1,\cdots,4$ on the left side
with $1',\cdots,4'$ on the right side yields again $Z$.
We now investigate the reduced density matrix $\bar{\rho}_{\bar
  A}=\mbox{Tr}_{\bar B}\bar\rho$ obtained by taking a partial trace
in auxiliary space.
$\bar{\rho}_{\bar A}$ is then a $4\times 4$ matrix whose eigenvalues
can be easily calculated.
Note that regions $\bar A,\bar B$ always have to contain an even
number of $\tau_{R,L}$-plaquettes so that the number of right and left
shift operators is the same. A discretization using more than four
plaquettes would, in the simple case considered, just add additional
zero eigenvalues. The entanglement measure $\bar S_{\rm ent}(\bar
A)=-\Tr\bar\rho_{\bar A}\ln\bar\rho_{\bar A}=\bar S_{\rm ent}(\bar B)$
is thus well-defined with $\lim_{T\to 0}\bar S_{\rm ent}= 2\ln 2$ and
$\bar S_{\rm ent}\sim 3(1+6\ln 2+2\ln T)/64T^2$ for $T\to\infty$, see
Fig.~\ref{Fig1}(e).

\begin{figure}[!ht]
\begin{center}
\includegraphics*[width=0.99\columnwidth]{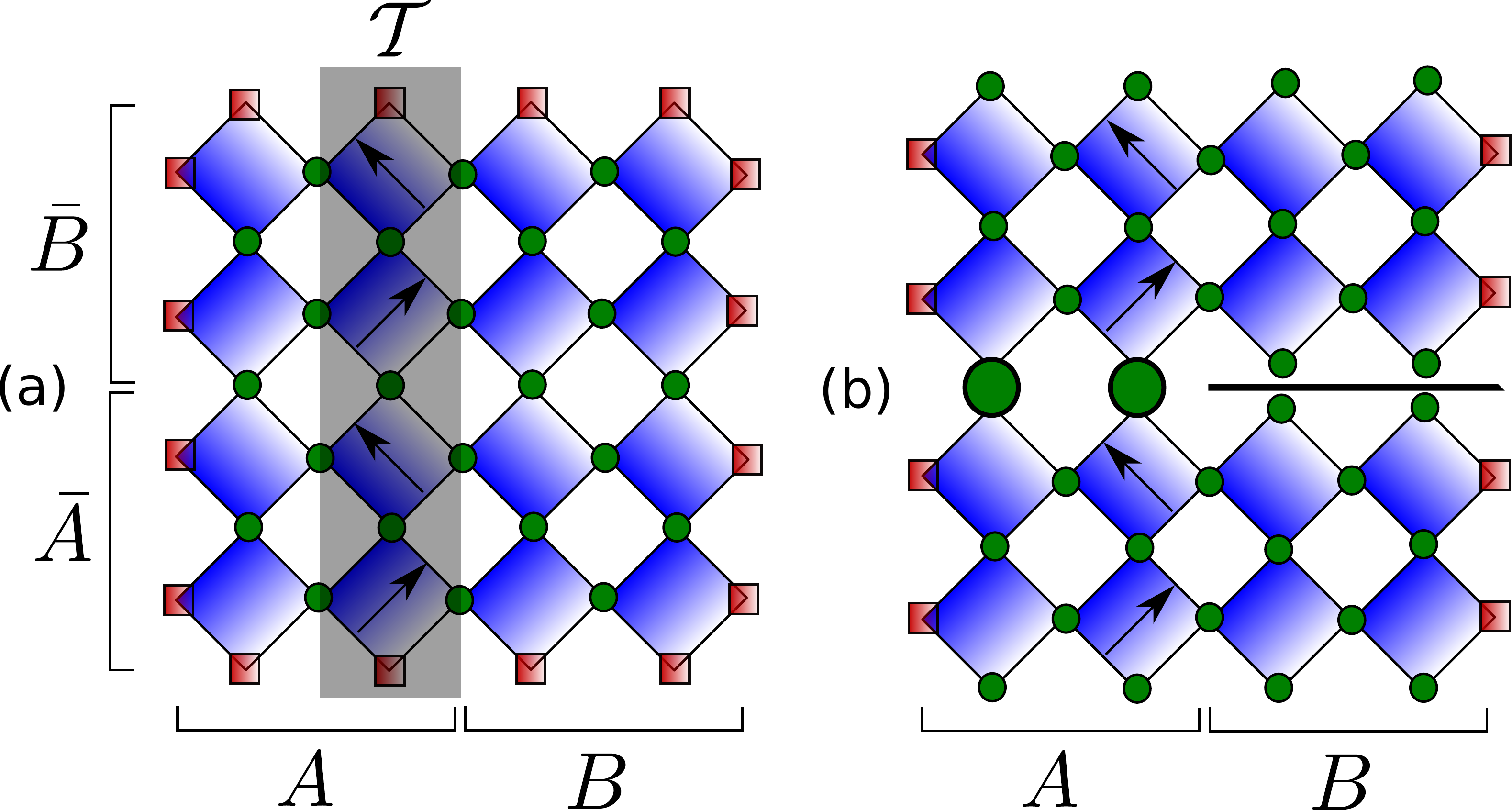}
\end{center}
\caption{(Color online) (a) Two-dimensional lattice obtained after the
  quantum-to-classical mapping. The shaded part is the QTM
  $\mathcal{T}$. Performing the trace over $\bar A$, $\bar B$ and the
  partial spatial trace over region $B$ yields $\rho_A$ while tracing
  over $A,B$ and taking the partial trace over region $\bar B$ yields
  $\bar\rho_{\bar A}$. (b) Geometry to calculate $\Tr\rho_A^2$.  The
  right part consists of two small cylinders while the left part is
  one big cylinder.}
\label{Fig3}
\end{figure}

Next, we want to generalize these considerations to one-dimensional
quantum systems. 
A Hamiltonian with short-range interactions can always be written as
$H=\sum_j h_{j,j+1}$, possibly in an enlarged unit cell. By using a
Trotter-Suzuki decomposition, the system can be mapped onto a
two-dimensional classical system in much the same way we have mapped
the qubits onto a one-dimensional classical system. A pictorial
representation is shown in Fig.~\ref{Fig3}(a) where each plaquette is
given by $\tau_{R,L}=T_{R,L}\exp(-\delta\beta h_{j,j+1})$.
Note that $[h_{j,j+1},h_{k,k+1}]\neq 0$ in general so that the mapping
induces an error which is of order $(\delta\beta)^2$ for the partition
function.


The canonical density matrix $\rho_{\rm c}$ for periodic boundary
conditions is shown pictorially in Fig.~\ref{Fig3}(a) with the l.h.s.
and r.h.s. indices traced over. Tracing instead over the upper and
lower indices and leaving the indices on the l.h.s. and r.h.s. open we
obtain the density matrix $\bar\rho =\mathcal{T}^L/Z$ where
$\mathcal{T}$ is a {\it quantum transfer matrix} (QTM) acting in
auxiliary space. The partition function is given by
$Z=\Tr_{A,B}\rho_c=\Tr_{\bar A,\bar B}\bar\rho$. For any $T>0$ the
thermodynamic limit can now be performed exactly with
$\lim_{L\to\infty} \bar\rho =|\Psi_R\rangle\langle\Psi_L|$. Here
$|\Psi_R\rangle$ and $\langle\Psi_L|$ are the left and right
eigenvectors belonging to the largest eigenvalue of $\mathcal{T}$ with
$\langle\Psi_L|\Psi_R\rangle=1$. This property can be easily
understood physically because all correlation lengths---determined by
the logarithm of the ratio of the leading to subleading eigenvalues of
$\mathcal{T}$---stay finite. $\bar\rho$ thus becomes a projector onto
the ground state in auxiliary space. Note that in this representation
$\bar\rho$ is non-symmetric. A symmetric representation is also
possible but would require a wider column transfer matrix
$\mathcal{T}$.  The reduced density matrix $\bar\rho_{\bar
  A}=\Tr_{\bar B} \bar\rho$ is then obtained by tracing over half of
the indices along the imaginary time axis.  Crucially, it is again
easy to show using a Schmidt decomposition and choosing biorthonormal
sets of right and left basis vectors that $\bar{S}_{\rm ent}\equiv
\bar{S}_{\rm ent}(\bar A)=\bar S_{\rm ent}(\bar B)$. $\bar\rho$ being
a projector thus guarantees that $\bar S_{\rm ent}$ is non-extensive
for all temperatures $T$.
It is thus quite natural to replace the projector $\rho$ onto the
ground state, considered for zero temperature, with the projector in
auxiliary space, $\bar\rho$, for an infinite one-dimensional system at
finite temperatures.

Because $\rho_A$ is not a projector for $T>0$, $S_{\rm ent}(T)$ is
difficult to calculate. Instead, the replica trick is often used
\cite{CalabreseCardy} which leads to the definition of Renyi entropies
$S_n(A)=\ln[\Tr(\rho^n_A)]/(1-n)$. A quantum-to-classical mapping
yields the geometry for $\Tr(\rho^2_A)$ shown in Fig.~\ref{Fig3}(b).
Separate QTM's for the right and left part of the system can then be
defined, however, an evaluation of $S_2(A)$ requires one to calculate
overlaps between the eigenstates of the QTM's \cite{WilmsTroyer} and
is thus much harder than obtaining $\bar S_{\rm ent}$. In the
following, we study in more detail the scaling properties of $\bar
S_{\rm ent}(T)$ and the spectra of $\bar\rho_{\bar A}$ for a symmetric
cut $\bar A=\bar B=\beta/2$.
\paragraph{Transverse Ising model}
We start with the transverse Ising model
\begin{equation}
\label{TrIs}
H=\sum_j \left(\sigma^z_j\sigma^z_{j+1}+\lambda\sigma^x_j\right)
\end{equation}
where $\vec{\sigma}$ are the Pauli matrices. This model shows a second
order phase transition at $\lambda=1$ \cite{Sachdev}.
To study $\bar S_{\rm ent}(T)$ and the entanglement spectra of
$\bar\rho_{\bar A}$ we use a transfer matrix renormalization group
(TMRG) algorithm
\cite{BursillXiang,WangXiang,Shibata,SirkerKluemperEPL}.
\begin{figure}[!ht]
\begin{center}
\includegraphics*[width=0.9\columnwidth]{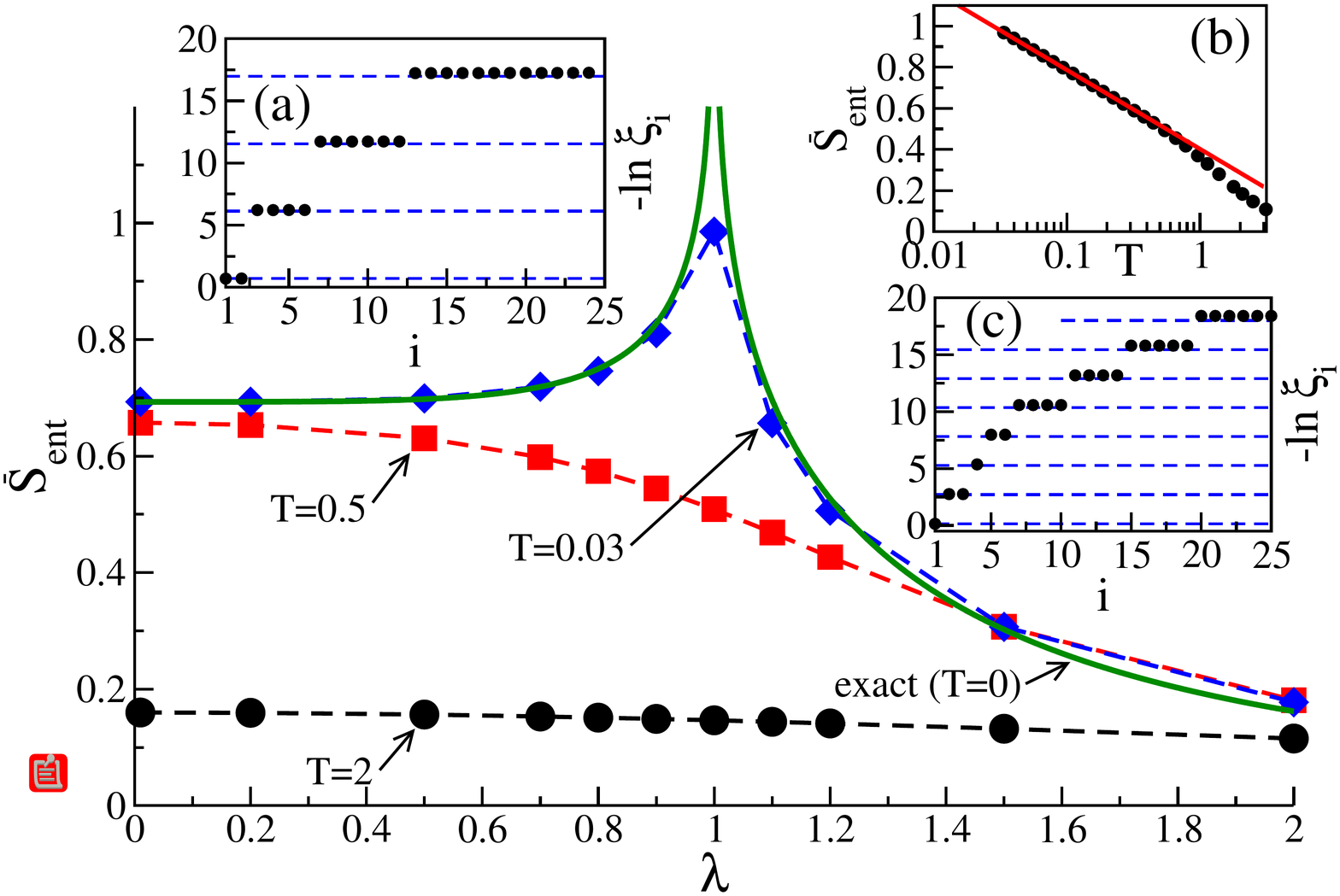}
\end{center}
\caption{(Color online) TMRG results are denoted by symbols. Main:
  $\bar S_{\rm ent}$ for various $T$ and exact result for $T=0$ (solid
  line). Eigenvalues $\xi_i$ of $\bar\rho_{\bar A}$ for (a)
  $\lambda=0.8$ and (c) $\lambda=1.2$ at $T=0.02$. The free fermion
  levels are denoted by dashed lines. (b) $\bar S_{\rm ent}(T)$ for
  $\lambda=1$ and low-temperature fit $S_{\rm ent}=-0.167\ln T
  +0.402$.}
\label{Ising}
\end{figure}
The entanglement spectra in the low-temperature limit are shown
exemplarily in Fig.~\ref{Ising}(a,b) and are equally spaced. The
degeneracies of the dominant eigenvalues are given by
$2,4,6,12,\cdots$ for $\lambda<1$ and $1,2,1,2,4,4,5,6,\cdots$ for
$\lambda>1$. As for the reduced density matrix $\rho_A$ obtained by a
spatial cut \cite{PeschelKaulke} these spectra can be explained by
using corner transfer matrices (CTM's). We find $\bar\rho_{\bar
  A}=\e^{-H_{CTM}}/\Tr\e^{-H_{CTM}}$ where $H_{CTM}=\sum\epsilon_j
n_j$ is a free fermion Hamilton operator.  For $\lambda>1$ all single
particle levels are twofold degenerate, $n_j=0,1,2$, and given by
$\epsilon_j=(2j+1)\pi K(\sqrt{1-1/\lambda^2})/K(1/\lambda)$ where
$K(x)$ is the complete elliptic integral of the first kind. For
$\lambda<1$ we have $\epsilon_j=j\pi
K(\sqrt{1-\lambda^2})/K(\lambda)$ where $\epsilon_0$ is
non-degenerate and all other levels twofold degenerate. 
Knowing the spectrum in the zero temperature
limit it is easy to calculate $\bar{S}_{\rm ent}(T\to 0)$
\cite{CalabreseCardy}. The result is shown as solid line in
Fig.~\ref{Ising}.
Right at the critical point the regular entanglement entropy for
an interval of length $\ell$ in an infinite chain shows a logarithmic
divergence with system size, $S_{\rm ent}=\frac{c}{3}\ln \ell +C_1$, with
central charge $c=1/2$ and a non-universal constant $C_1$
\cite{CalabreseCardy}.  From this result 
it follows immediately by a conformal mapping that
\begin{equation}
\label{logscaling}
\bar S_{\rm ent}(T) =\frac{c}{3}\ln(v/T) +C_1
\end{equation}
with $v$ being the velocity of the elementary excitations. This result
is universal for critical systems.  For the transverse Ising model it
is in excellent agreement with a two-parameter fit of the numerical
data, see Fig.~\ref{Ising}(b).
\paragraph{$XXZ$ model and boundary locality}
As a further example, we calculate $\bar S_{\rm ent}$ and
$\bar\rho_{\bar A}$ for the $XXZ$ model
\begin{equation}
\label{XXZ}
H=J\sum_j \left\{S^x_jS^x_{j+1}+S^y_jS^y_{j+1}+\Delta S^z_jS^z_{j+1}\right\}\, .
\end{equation}
Here $\vec{S}$ is a spin-$1/2$ operator and $\Delta$ parametrizes the
exchange anisotropy. The model is critical for $-1<\Delta\leq 1$ with
central charge $c=1$. As shown in Fig.~\ref{Fig4} for three different
values of $\Delta$, Eq.~(\ref{logscaling}) does indeed describe $\bar
S_{\rm ent}$ for $T\ll J$. 
\begin{figure}[!ht]
\begin{center}
\includegraphics*[width=0.8\columnwidth]{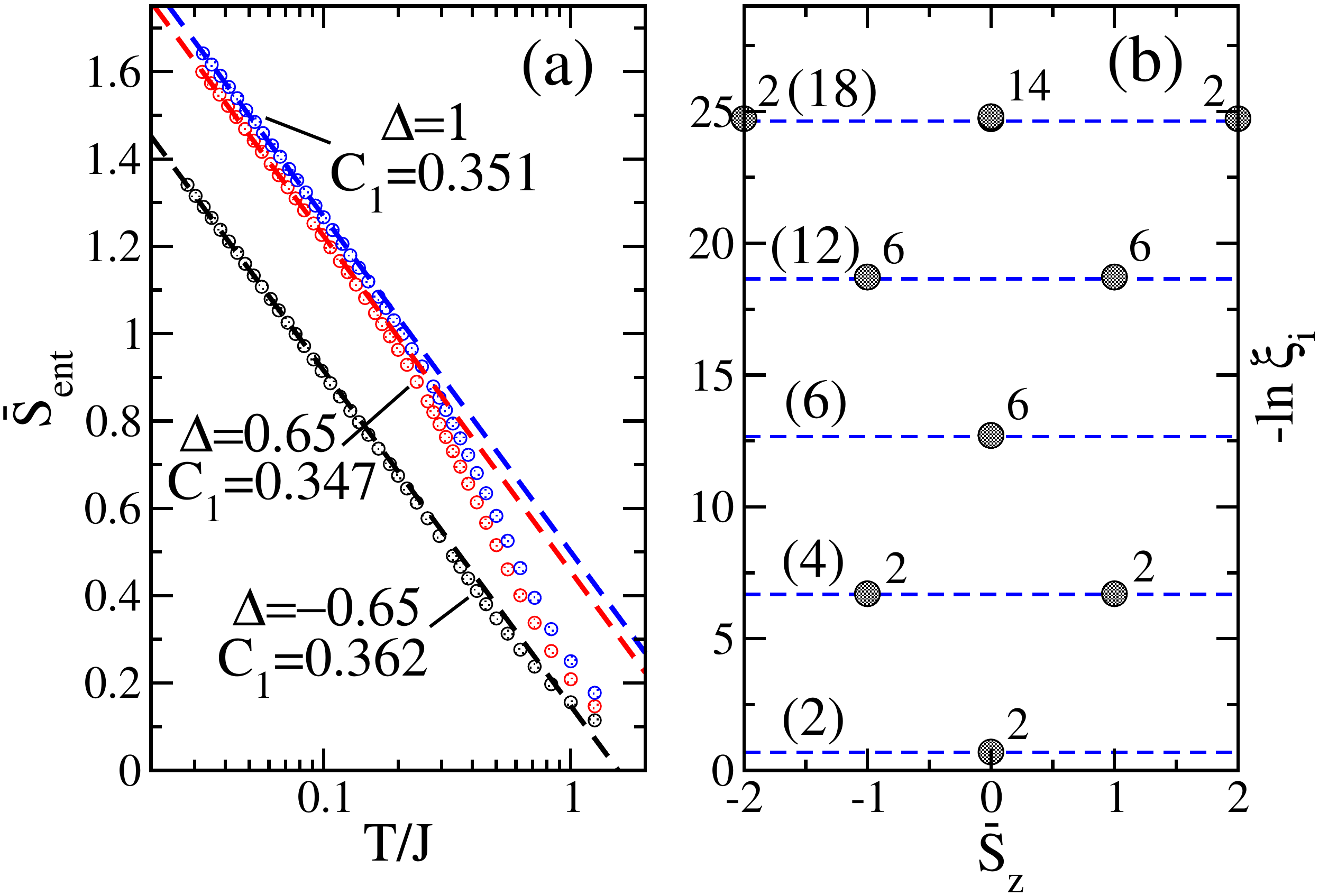}
\end{center}
\caption{(Color online) (a) One-parameter fits (lines) for $T/J\leq
  0.1$ of TMRG data (symbols) according to (\ref{logscaling}) with
  $c=1$, $v$ as known from the Bethe ansatz, and $C_1$ as indicated.
  (b) Spectrum of $\bar\rho_{\bar A}$ for $\Delta=10$ and $T/J=0.02$.
  The numbers give the multiplicities of each level (total in
  brackets), the dashed lines mark the free fermion levels.}
\label{Fig4}
\end{figure}
The spectrum of $\rho_A$ in the non-critical case $\Delta>1$ can be
constructed from a free-fermion Hamiltonian \cite{PeschelKaulke} and
we find again that the same is also true for $\bar\rho_{\bar A}$ at
low temperatures. As an example, the spectrum for $\Delta=10$ is shown
in \ref{Fig4}(b). The degeneracies of the free Fermion levels are the
same as for the transverse Ising model with $\lambda<1$, however, now
$\epsilon_j=2j\cosh^{-1}\Delta$ \cite{PeschelKaulke}. $S^z_{\rm
  tot}=\sum_j S^z_j$ commutes with the Hamiltonian and is thus a good
quantum number. For the QTM $\mathcal{T}$ it follows that
$\bar{S}^z_{\rm R,L}=\sum_k (-1)^k \bar (S_k^z)_{R,L}$ is a good
quantum number where $(\bar S_k^z)_{R,L}$ acts at site $k$ in
auxiliary space on the r.h.s.~respectively l.h.s.~of $\mathcal{T}$.
Matrix elements of $\mathcal{T}$ can thus only be non-zero if
$\bar{S}^z\equiv\bar{S}^z_{\rm R} = \bar{S}^z_{\rm L}$ and eigenvalues
of $\bar\rho_{\bar A}$ can be classified according to $\bar{S}^z$, see
Fig.~\ref{Fig4}(b). For $\rho_A$ it has been shown that its spectrum
as a function of $S^z_{\rm tot}$ can be constructed for $\Delta\gg 1$
by local perturbation theory in the spin exchange operator
\cite{AlbaHaque}. The same local perturbation theory can also be
performed for the QTM $\mathcal{T}$. Keeping in mind that we have
periodic boundary conditions along the imaginary time direction so
that the partial trace introduces two boundaries, the degeneracies of
each level for $\bar\rho_{\bar A}$ as a function of $\bar S^z$ are
exactly the same as for $\rho_A$ with two boundaries classified by
$S^z_{\rm tot}$. This demonstrates explicitly that a boundary law also
holds for $\bar\rho_{\bar A}$ obtained by a cut along the imaginary
time direction.
\paragraph{Ladders and two-dimensional models}
The quantum-to-classical mapping of a finite two-dimensional system of
extent $L_1\times L_2$ leads to a three-dimensional system with
dimensions $\beta\times L_1\times L_2$. In order to calculate the
mutual information $\mathcal{I}_{A,B}$ following a spatial cut one can
again use the replica trick. In particular, the Renyi entropy $S_2$
has been obtained from the three-dimensional generalization of the
geometry shown in Fig.~\ref{Fig3}(b) by quantum Monte Carlo (QMC)
simulations \cite{MelkoKallin,IsakovHastings,SinghHastings}.  However,
the non-trivial geometry makes these calculations rather complicated
and $S_{\rm ent}$ remains inaccessible. On the contrary, the numerical
calculation of $\bar S_{\rm ent}$ is straightforward also in two
dimensions.
For $L_1$ and $L_2$ both finite, one can use the mutual information
$\bar{\mathcal{I}}_{\bar A, \bar B} =\bar S_{\rm ent}(\bar A) +\bar
S_{\rm ent}(\bar B) +\Tr\bar\rho\ln\bar\rho$ to detect finite
temperature phase transitions. Furthermore, one can perform the
thermodynamic limit in one of the spatial dimensions with
$\lim_{L_1\to\infty}\bar\rho(L_1,L_2,\beta)=|\Psi_R\rangle\langle\Psi_L|$
becoming again a projector so that $\lim_{T\to\infty}\bar S_{\rm
  ent}=0$ and a subtraction of a 'thermal part' is no longer required.

\paragraph{Discussion}
We have analyzed the entanglement entropy $\bar S_{\rm ent}$ of a
reduced density matrix obtained after a quantum-to-classical mapping
and a partial trace in imaginary time direction. Our results show that
transfer matrix DMRG algorithms are as efficient in simulating
one-dimensional quantum systems in the thermodynamic limit at finite
temperatures as regular DMRG algorithms are for finite systems at
$T=0$. For two- and three-dimensional systems $\bar S_{\rm ent}$ can
be used to investigate finite temperature phase transitions. For QMC,
in particular, this might be a viable and---due to the simpler
geometry---easier approach than calculating Renyi entropies.


\acknowledgments I want to thank P. Calabrese, J. Cardy,
R.~Dillenschneider, M. Haque, and, in particular, I. Peschel for
helpful discussions. I thank the Galileo Galilei Institute for
Theoretical Physics for their hospitality and the INFN for partial
support during completion of this work. I also acknowledge support by
the DFG via the SFB/TR 49 and by the graduate school of excellence
MAINZ.


\end{document}